\documentclass[twoside,english]{emulateapj}
\usepackage{graphicx}

\makeatletter

\providecommand{\tabularnewline}{\\}

\usepackage{times}

\makeatother

\begin{document}

\title{The efficiency of resonant relaxation around a massive black hole}

\author{Ehud Eilon\altaffilmark{1}, G\'abor Kupi and Tal Alexander\altaffilmark{2}}

\affil{Weizmann Institute of Science, Rehovot 76100, Israel}

\altaffiltext{1}{Present address: Israel Aerospace Industries, MLM Division, PO Box 45, Beer-Yaakov, 70350, Israel}

\altaffiltext{2}{William Z. and Eda Bess Novick Career Development Chair}

\begin{abstract}
Resonant relaxation (RR) is a rapid relaxation process that operates
in the nearly-Keplerian potential near a massive black hole (MBH).
RR dominates the dynamics of compact remnants that inspiral into a
MBH and emit gravitational waves (extreme mass ratio inspiral events,
EMRIs). RR can either increase the EMRI rate, or strongly suppress
it, depending on its still poorly-determined efficiency. We use small-scale
Newtonian $N$-body simulations to measure the RR efficiency and to
explore its possible dependence on the stellar number density profile
around the MBH, and the mass-ratio between the MBH and a star (a single-mass
stellar population is assumed). We develop an efficient and robust
procedure for detecting and measuring RR in $N$-body simulations.
We present a suite of simulations with a range of stellar density
profiles and mass-ratios, and measure the mean RR efficiency in the
near-Keplerian limit. We do not find a statistically significant dependence
on the density profile or the mass-ratio. Our numerical determination
of the RR efficiency in the Newtonian, single-mass population approximations,
suggests that RR will likely \emph{enhance} the EMRI rate by a factor
of a few over the rates predicted assuming only slow stochastic two-body
relaxation. 
\end{abstract}


\section{Introduction}

Dynamical relaxation processes near massive black holes (MBH) in galactic
centers affect the rates of strong stellar interactions with the MBH,
such as tidal disruption, tidal dissipation, or gravitational wave
(GW) emission \citep[e.g.][]{ale05}. These relaxation processes may
also be reflected by the dynamical properties of the different stellar
populations there \citep{hop+06a}, as observed in the Galactic Center
\citep{gen+00,pau+06}. Of particular importance, in anticipation
of the planned Laser Interferometer Space Antenna (LISA) GW detector,
is to understand the role of relaxation in regulating the rate of
GW emission events from compact remnants undergoing quasi-periodic
extreme mass ratio inspiral (EMRI) into MBHs. 

Two-body relaxation, or non-coherent relaxation (NR), is inherent
to any discrete large-N system, due to the cumulative effect of uncorrelated
two-body encounters. These cause the orbital energy $E$ and the angular
momentum $J$ to change in a random-walk fashion ($\propto\!\sqrt{t}$)
on the typically long NR timescale $T_{\mathrm{NR}}$. In contrast,
when the gravitational potential has approximate symmetries that restrict
orbital evolution (e.g. fixed ellipses in a Keplerian potential; fixed
orbital planes in a spherical potential), the perturbations on a test
star are no longer random, but correlated, leading to coherent ($\propto\! t$)
torquing of $J$ on short timescales, while the symmetries hold. Over
longer times, this results in resonant relaxation (RR) (\citealt{rau+96,rau+98};
 \S \ref{ss:RR}), a rapid random walk of $J$ on the typically short
RR timescale $T_{\mathrm{RR}}\!\ll\! T_{\mathrm{NR}}$. RR in a near-Keplerian
potential can change both the direction and magnitude of $\mathbf{J}$
({}``scalar RR''), thereby driving stars to near-radial orbits that
interact strongly with the MBH. RR in a near-spherical potential can
only change the direction of $\mathbf{J}$ ({}``vector RR'').

RR is particularly relevant in the potential near a MBH, where compact
EMRI candidates originate. \citet{hop+06a} show that RR dominates
EMRI source dynamics. Depending on its still poorly-determined efficiency,
RR can either increase the EMRI rate over that predicted assuming
NR only, or if too efficient, it can strongly suppress the EMRI rate
by throwing the compact remnants into infall (plunge) orbits (cf Fig.
\ref{f:EMRI} below) that emit a single, non-periodic and hard to
detect GW burst. A prime motivation for the systematic numerical investigation
of RR efficiency presented here, are the still open questions about
the implications of RR for EMRI rates and orbital properties.

This paper is organized as follows. In \S \ref{s:theory} we briefly
review the theory of NR and RR relaxation and derive a new relation
between scalar and vector RR. In \S \ref{s:ACF} we describe our
method of analyzing and quantifying the effects of RR in $N$-body
simulations, which are described in \S \ref{s:simulations}. We present
our results in \S \ref{s:results} and discuss and summarize them
in \S \ref{s:conclusions}.

\section{Theory}

\label{s:theory}

\subsection{Non-coherent Relaxation (NR)}

\label{ss:NR}

The NR time for $E$-relaxation, $T_{\mathrm{NR}}^{E}$, corresponds
to the time it takes non-coherent 2-body interactions to change the
stellar orbital energy by order of itself, $\left|\Delta E\right|\!\sim\! E$
(by stellar dynamical definition convention, $E\!>\!0$ for a bound
orbit). Similarly, the NR time for $J$-relaxation, $T_{\mathrm{NR}}^{J}$,
corresponds to the time it takes the stellar orbital angular momentum
to change by order of the circular angular momentum $\left|\Delta J\right|\!\sim\! J_{c}$,
where near a MBH of mass $M$, $J_{c}\!=\! GM/\sqrt{2E}$. The $E$-relaxation
timescale can be estimated by considering the rate $\Gamma$ of gravitational
collisions in a system of size $R$ at a relative velocity $v$, between
a test star and $N$ field stars of mass $m$ and space density $n\!\sim\! N/R^{3}$,
at the minimal impact parameter where the small angle deflection assumption
still holds, $r_{min}\!\sim\! Gm/v^{2}$. The collision rate is then
$\Gamma\!\sim\! nvr_{min}^{2}\!\sim\! G^{2}m^{2}n/v^{3}$. Taking
into account also collisions at larger impact parameters increases
the rate by the Coulomb logarithm factor $\textrm{ln}\Lambda\!\sim\!\textrm{ln}(R/r_{min})$.
Therefore, $T_{\mathrm{NR}}^{E}\!\sim\! v^{3}/(G^{2}m^{2}n\textrm{ln}\Lambda)$.
Near the MBH $v^{2}\!\sim\! GM/R$, and so $\ln\Lambda\!\sim\!\ln Q$,
where $Q\!\equiv\! M/m$ is the mass ratio.

When the stars move under the influence of the central MBH ($Q\!\gg\! N$),
the relaxation time can be expressed as $T_{\mathrm{NR}}^{E}\!\sim\!(M/m)^{2}P/(N\textrm{ln}\Lambda)$,
where $P\!=\!2\pi\sqrt{R^{3}/GM}$ is the Keplerian period. Following
the notation of \citet{rau+96} (RT96), the NR changes in $E$, $J$
and $\mathbf{J}$ over the dimensionless time-lag $\tau\!\equiv\!(t_{2}-t_{1})/P_{1}$
are\begin{eqnarray}
\left|\Delta E\right|/E & \equiv & \left|E_{2}-E_{1}\right|/E_{1}=\alpha_{\Lambda}\sqrt{N}(m/M)\sqrt{\tau}\,,\label{e:alphaL}\\
\left|\Delta J\right|/J_{c} & \equiv & \left|J_{2}-J_{1}\right|/J_{c,1}=\eta_{s\Lambda}\sqrt{N}(m/M)\sqrt{\tau}\,,\label{e:etasL}\\
\left|\Delta\mathbf{J}\right|/J_{c} & \equiv & \left|\mathbf{J}_{2}-\mathbf{J}_{1}\right|/J_{c,1}=\eta_{v\Lambda}\sqrt{N}(m/M)\sqrt{\tau}\,,\label{e:etavL}\end{eqnarray}
where $\alpha_{\Lambda}\!\equiv\!\alpha\sqrt{\textrm{ln}\Lambda}$
and $\eta_{s,v\Lambda}\!\equiv\!\eta_{s,v}\sqrt{\textrm{ln}\Lambda}$
are dimensionless constants, whose exact values are system-dependent
and cannot be estimated with accuracy without detailed calculations
or simulations. The corresponding NR timescales are related to these
coefficients by $T_{\mathrm{NR}}^{E}\!=\!(M/m)^{2}P/(N\alpha_{\Lambda}^{2})$,
$T_{\mathrm{NR}}^{J}\!=\!(M/m)^{2}P/(N\eta_{s\Lambda}^{2})$ and $T_{\mathrm{NR}}^{\mathbf{J}}\!=\!(M/m)^{2}P/(N\eta_{v\Lambda}^{2})$.

\subsection{Resonant Relaxation (RR)}

\label{ss:RR}

When the potential has symmetries that restrict the orbital evolution,
for example to fixed ellipses in the potential of a point mass, or
to planar annuli in a spherical potential, the perturbations on a
test star are no longer random, but correlated. This leads to a coherent
changes in $\mathbf{J}$ on times $P\!\ll\! t\!<\! t_{\omega}$, $\Delta\mathbf{J}\!=\!\mathbf{T}t$,
by the residual torque $\left|\mathbf{T}\right|\!\sim\!\sqrt{N}Gm/R$
exerted by the $N$ randomly oriented, orbit-averaged mass distributions
of the surrounding stars (mass wires for elliptical orbits in a Kepler
potential, mass annuli for rosette-like orbits in a spherical potential).
The coherence time $t_{\omega}$ is set by deviations from the true
symmetry, which lead to a gradual orbital drift and to the randomization
of $\mathbf{T}$. For example, the enclosed stellar mass leads to
non-Keplerian retrograde precession; General Relativity leads to prograde
precession. Ultimately, the coherent torques themselves randomize
the orbits (alternatively, this can be viewed as the result of potential
fluctuations due to the finite number of stars). The effective coherence
time is set by the shortest de-coherence (quenching) process in the
system. The accumulated change over $t_{\omega}$, $\left|\Delta\mathbf{J}_{\omega}\right|\!\sim\!\left|\mathbf{T}t_{\omega}\right|$,
then becomes the basic step-size, or mean free path in $\mathbf{J}$-space,
for the long-term ($t\!\gg\! t_{\omega}$) non-coherent ($\propto\!\sqrt{t}$)
relaxation of $\mathbf{J}$. Since this step-size is large, RR can
be much faster than NR. The RR timescale $T_{\mathrm{RR}}$ is then
defined by $\Delta J/J_{c}\!=\!(\Delta J_{\omega}/J_{c})\sqrt{t/t_{\omega}}\!\equiv\!\sqrt{t/T_{\mathrm{RR}}}$.
Note that the relaxation of \emph{$E$} is not affected by RR because
the potential of the system is stationary on the coherence timescale,
and so $E$ changes incoherently on all time scales. The torques exerted
by elliptical mass wires in a Kepler potential can change both the
direction and magnitude of \textbf{J}. In contrast, the torques exerted
by planar annuli can only change the direction of \textbf{J}.

Here we consider only Newtonian dynamics. The coherence timescale
for scalar RR is determined by the time it takes for the orbital apsis
to precess by angle $\sim\!\pi$ due to the potential of the enclosed
stellar mass ({}``mass precession''),\begin{equation}
t_{\omega}=t_{M}=A_{M}\left(M\left/Nm\right.\right)P\,,\end{equation}
where $A_{M}$ is an $O(1)$ factor reflecting the approximations
in this estimate. The coherence timescale for vector RR is determined
by the time it takes for the coherent torques to change $\Delta\left|\mathbf{J}\right|\sim J_{c}$
(alternatively, this is the timescale $t_{\phi}=(\phi/\Delta\phi_{\star})P/2$
to accumulate $O(1)$ fluctuations in the stellar potential $\Delta\phi_{\star}\!\sim\!\sqrt{N}Gm/R$
relative to the total gravitational potential $\phi$ as the stars
rotate by $\sim\!\pi$ on their orbits), 

\begin{equation}
t_{\omega}=t_{\phi}=A_{\phi}\left(\sqrt{N}\left/2\mu\right.\right)P\simeq\left[A_{\phi}(M/m)\left/2\sqrt{N}\right.\right]P\,,\end{equation}
where $A_{\phi}$ is an $O(1)$ factor, $\mu\!=\! Nm/(M+Nm)$ and
where the approximate equality is for the Keplerian limit $Nm\!\ll\! M$.
Following RT96, the RR changes in $J$ and $\mathbf{J}$ during the
coherent phase $(\tau\!<\!\tau_{\omega})$ can be expressed as

\begin{eqnarray}
\left|\Delta J\right|/J_{c} & \equiv & \left|J_{2}-J_{1}\right|/J_{c,1}=\beta_{s}\sqrt{N}(m/M)\tau\,,\label{e:betas}\\
\left|\Delta\mathbf{J}\right|/J_{c} & \equiv & \left|\mathbf{J}_{2}-\mathbf{J}_{1}\right|/J_{c,1}=\beta_{v}\sqrt{N}(m/M)\tau\,,\label{e:betav}\end{eqnarray}
where the $O(1)$ dimensionless coefficients $\beta_{s}$ and $\beta_{v}$
depend on the parameters of the system and reflect the uncertainties
introduced by the various approximations and simplification of this
analysis. Accurate determination of their values requires detailed
calculations or simulations. 

The scalar RR change $\Delta J$ on time-lags $\tau\!\gg\!\tau_{M}$
is then 

\begin{equation}
\left|\Delta J\right|/J_{c}\!\equiv\!\left|J_{2}\!-\! J_{1}\right|/J_{c,1}\!=\!\beta_{s}\sqrt{A_{M}(m/M)}\sqrt{\tau}\,,\end{equation}
and the scalar RR timescale is 

\begin{equation}
T_{\mathrm{RR}}^{J}=\left[(M/m)\left/A_{M}\beta_{s}^{2}\right.\right]P\,.\end{equation}
The RR efficiency factor $\chi\!=\!(\beta_{s}/\beta_{s,\mathrm{RT}96})^{2}$
defined by \citet{hop+06a} expresses how much shorter $T_{\mathrm{RR}}^{J}$
is relative to the value estimated by RT96. Scalar RR is faster than
NR by a factor $T_{\mathrm{NR}}^{J}/T_{\mathrm{RR}}^{J}\!\propto\!(M/m)/N\ln\Lambda$.
Similarly, the vector RR change $\left|\Delta\mathbf{J}\right|$ on
time-lags $\tau\!\gg\!\tau_{\phi}$ in the Keplerian limit is 

\begin{equation}
\left|\Delta\mathbf{J}\right|/J_{c}\!\equiv\!\left|\mathbf{J}_{2}\!-\!\mathbf{J}_{1}\right|/J_{c,1}\!=\!\beta_{v}\sqrt{\frac{1}{2}A_{\phi}\sqrt{N}(m/M)}\sqrt{\tau}\,,\end{equation}
and the vector RR timescale is

\begin{equation}
T_{\mathrm{RR}}^{\mathbf{J}}=\left[2(M/m)\left/\sqrt{N}A_{\phi}\beta_{v}^{2}\right.\right]P\,.\end{equation}

RT96 performed a limited set of near-Keplerian simulations to check
their predictions. They analyzed the results both star by star and
in the average and observed the coherent growth of $\Delta J/J_{c}$
and $\left|\Delta\mathbf{J}\right|/J_{c}$ relative to the simulation's
initial values. Although the evolution of the of these quantities
for any single star was very noisy and the proportionality factors
had a very large scatter, RR was clearly observed, as predicted.

\subsubsection{Relation between scalar and vector RR}

\label{sss:sv}

The population averages of the scalar and vector coefficients $\eta_{s,v}$
and $\beta_{s,v}$, are not independent quantities. Rather, $\left\langle \eta_{v}\right\rangle \!=\! c\left\langle \eta_{s}\right\rangle $,
and $\left\langle \beta_{v}\right\rangle \!=\! c\left\langle \beta_{s}\right\rangle $
(on timescales $t\!<\! t_{\omega}$), where the constant $c$ depends
on the averaging procedure. Here we average $\left|\Delta\mathbf{J}\right|(\tau)$
and $\Delta J(\tau)$ by the rms over the stellar population (\S
\ref{s:ACF}). We focus on the limit where $\left|\Delta\mathbf{J}\right|/J\!\ll\!1$,
and assume that the change is isotropic on average, $\left\langle \mathbf{\left|\Delta J\right|}^{2}\right\rangle \!=\!3\left\langle \left|\Delta J_{i}\right|^{2}\right\rangle \!=\!\delta^{2}$,
(\emph{$i\!=\! x,y,z$}), as is indicated by our simulations (\S
\ref{s:simulations}). The rms of vector RR is $\delta$. Defining
the $z$-axis along $\mathbf{J}$, then $\Delta J\!=\!\Delta J_{z}$
in the small change limit, and its rms is $\delta/\sqrt{3}$. Therefore
(see Eqs. \ref{e:etasL}, \ref{e:etavL}, \ref{e:betas}, \ref{e:betav})
$\eta_{v}\!=\!\sqrt{3}\eta_{v}$ for all $\tau$ and $\beta_{v}\!=\!\sqrt{3}\beta_{s}$
for $\tau\!\ll\!\tau_{M},\tau_{\phi}$, as is indeed seen in the simulations%
\footnote{It can be shown that if the population's mean absolute difference
$\left\langle \left|\Delta\mathbf{J}\right|\right\rangle $ is used
to define $\beta_{s,v}$, then $\beta_{v}\!=\!2\beta_{s}$.%
} (Fig. \ref{f:Q1e6}). 

The constant ratio between the population averages of the vector and
scalar coefficients is in part a geometrical effect (1D vs 3D changes)
and in part a reflection of the symmetries of the perturbations (isotropic
$\Delta\mathbf{J}$). This strict proportionality is not expected
to hold when $\left|\Delta\mathbf{J}\right|/J\!\sim\!1$, for example
when only very eccentric stars ($J\rightarrow0$) are included in
the sample, or for very long time-lags.

\section{RR detection by auto-correlation analysis}

\label{s:ACF}

Given the high computational cost of the $N$-body simulations, and
the very large variance in the evolution of individual orbits, it
is essential to extract the RR signal as efficiently and robustly
as possible from the simulated data. After some experimentation, we
adopted the auto-correlation analysis for detecting and measuring
RR in $N$-body simulation snapshots. Our procedure is as follows. 

(1) The stellar phase space coordinates are transformed to the rest-frame
of the MBH, which is almost identical with the center of mass in the
near-Keplerian system close to the MBH.

(2) The energy $(E_{i}^{(n)})$, angular momentum $(\mathbf{J}_{i}^{(n)})$,
circular angular momentum $(J_{c,i}^{(n)})$ and Keplerian period
$(P_{i}^{(n)})$ are calculated for the $n$'th star $(n\!=\!1\ldots N)$
at discrete times $t_{i}$ in the simulation ({}``snapshots''). 

(3) To make full use of the data, we assign a normalized time-lag
$\tau_{ji}^{(n)}\!=\!(t_{j}\!-\! t_{i})/P_{i}^{(n)}$ for each pair
of times $(t_{j}\!>\! t_{i})$. For each of the \emph{$N$} stars,
we calculate the normalized energy and angular momentum differences
at all lags, $(\Delta E/E)_{ji}^{(n)}\!=\!(E_{j}^{(n)}\!-\! E_{i}^{(n)})/E_{i}^{(n)}$,
$(\Delta J/J_{c})_{ji}^{(n)}\!=\!(J_{j}^{(n)}\!-\! J_{i}^{(n)})/J_{c,i}^{(n)}$
and $(\Delta\mathbf{J}/J_{c})_{ji}^{(n)}\!=\!\left|\mathbf{J}_{j}^{(n)}\!-\!\mathbf{J}_{i}^{(n)}\right|/J_{c,i}^{(n)}$. 

(4) The differences from all stars are binned into discrete $\tau$-bins
according to their associated $\tau_{ij}$. The bin rms values and
their standard deviations are calculated and plotted against the bin's
average time-lag $\tau\!=\!\left\langle \tau_{ij}\right\rangle $,
thereby creating the auto-correlation curve. 

By using all possible time-lags recorded in the simulation, this approach
makes maximal use of the entire data set and averages over the strongly
fluctuating individual relaxation curves (cf the RT96 procedure, \S
\ref{ss:RR}). However, this method is not entirely free of bin-to-bin
bias. Since the number of orbital periods completed by a star with
a mean period $P_{i}^{(n)}$ over the simulation time $t_{\mathrm{sim}}$
is $t_{\mathrm{sim}}/P^{(n)}\!=\!\tau_{max}^{(n)}$, long-period stars
will not contribute to a high-$\tau$ bin, $\left\langle \tau\right\rangle _{i}$
if $\tau_{max}^{(n)}\!<\!\left\langle \tau\right\rangle _{i}$. Conversely,
short-period stars will not contribute to a low-$\tau$ bin, $\left\langle \tau\right\rangle _{i}$,
if $\tau_{min}^{(n)}\!>\!\left\langle \tau\right\rangle _{i}$, where
$\tau_{min}^{(n)}\!=\!\min_{j}(t_{j+1}\!-\! t_{j})/P^{(n)}$ is set
by the minimal time difference between consecutive snapshots. In extreme
cases some stars may not contribute to the relaxation curve at any
$\left\langle \tau\right\rangle _{i}$. Since long and short period
stars could well have systematically different responses to RR, this
introduces bias to the relaxation curve. This bias can be minimized,
at the cost of losing some information, by using only the middle range
of the $\tau$ -bins, or at a computational cost, by longer simulations
with a higher snapshot rate.

\section{Simulations}

\label{s:simulations}

\begin{figure}
\noindent \begin{centering}
\includegraphics[width=1\columnwidth]{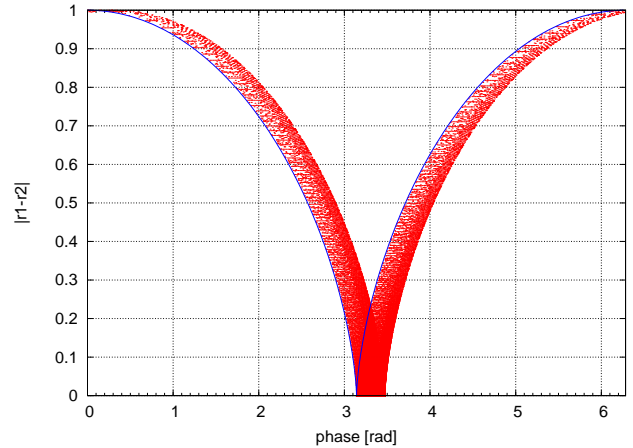}
\par\end{centering}

\caption{\label{f:phase}Phase drift in a 2-body system over a time of $\tau_{\mathrm{sim}}=1.35\times10^{4}$
for a $Q=3\times10^{-7}$ mass ratio and an extreme eccentricity of
$e=0.99995$. The line is the initial phase curve ($r$ as function
of $\psi$); the dots are the simulated data. The phase drift of $\Delta\psi=0.2875$
corresponds to $\tau_{\Delta\psi}\simeq3.3\times10^{5}$.}

\end{figure}

\begin{figure}[t]
\begin{centering}
\includegraphics[width=1\columnwidth]{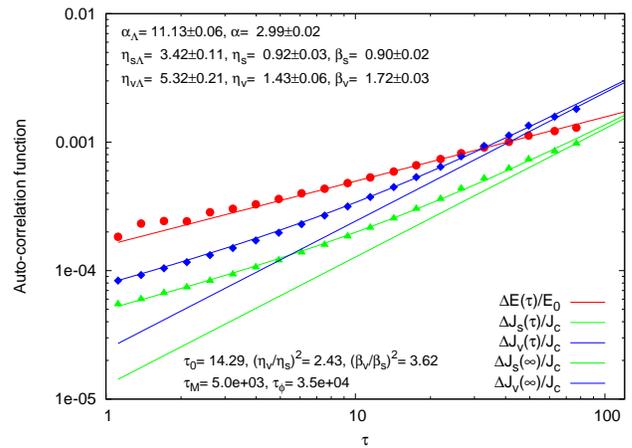}
\par\end{centering}

\caption{\label{f:evol}The measured (points) and fitted (lines) auto-correlation
curves for $\Delta E/E$, $\left|\Delta J\right|/J_{c}$ and $\left|\Delta\mathbf{J}\right|/J_{c}$
in a $Q\!=\!10^{6}$, $\gamma\!=\!1.5$ simulation with $N\!=\!200$
particles. The points are the results of the $N$-body simulation,
the thick lines are the predicted theoretical curves; the thin straight
lines show the asymptotic ($\tau\!\gg\!\tau_{0}$) linear behavior.
The best fit parameters and related quantities are also listed.}

\begin{centering}
\label{f:Q1e6}
\par\end{centering}
\end{figure}
Our $N$-body code uses a 5th order Runge-Kutta integrator with individual
time-steps and pair-wise K-S regularization \citep{kus+65}, without
gravity softening. The time steps were chosen to conserve total energy
at the level of $\Delta E_{\mathrm{tot}}/E_{\mathrm{tot}}\!\sim\! O(10^{-5})-O(10^{-8})$,
well below the NR energy changes expected in the simulations.

The $N$-body code must maintain phase coherence to a high enough
accuracy to reliably simulate RR. It is thus important to verify that
the numerical phase drift in the simulation is much smaller than that
due to physical processes, such as mass precession. Experience shows
that most of the phase drift $\Delta\psi$ accumulates near periapse,
where the acceleration is largest, and is therefore more pronounced
in eccentric orbits. We estimated the phase drift by integrating a
highly eccentric ($e\!=\!0.99995$) 2-body system over many orbits,
and plotted the relative distance between the 2 masses as function
of the orbital phase (Fig. \ref{f:phase}), as it evolved during the
simulation. The phase spread relative to the theoretical phase curve,
$\Delta\psi$, was measured near $\psi\!=\!\pi/2$. We define the
phase de-coherence timescale as $\tau_{\Delta\psi}\!\equiv\!(\pi/\Delta\psi)\tau_{\mathrm{sim}}$.
We estimate conservatively that our near-Keplerian simulation typically
have $\tau_{\Delta\psi}\!>\!10^{5}$ (This is for highly eccentric
orbits. We do not detect any phase drift up to $\tau\!\sim\!10^{5}$
in orbits with moderate eccentricities). Since $\tau_{\Delta\psi}\!>\!\tau_{w}\!\gg\!\tau_{\mathrm{sim}}$
for the models simulated here (cf Fig. \ref{f:evol}), it can be safely
neglected.

Our simulations consisted of 200 particles (including the MBH as a
free particle). The initial orbital semi-major axes were randomly
drawn from a $\rho(a)\mathrm{d}a\!\propto\! a^{2-\gamma}\mathrm{d}a$
distribution for $\gamma\!=\!1,\,1.5,\,1.75$, for $a$ in the range
$(0,1/2)$, with eccentricities drawn from a thermal $\rho(e)\mathrm{d}e=2e\mathrm{d}e$
distribution, with random phases, orbital orientations and isotropic
velocities. This distribution corresponds to an $r^{-\gamma}$ number
density distribution with an outer cutoff at radius $r\!=\! R\!=\!1$
from the MBH (in dimensionless units where $G\!=\!1$, $M\!+\! Nm\!=\!1$).
These stellar cusps span a wide range of possible physical scenarios
\citep[e.g.][]{bah+77}, and in particular those of LISA targets,
which are expected to be relaxed galactic nuclei \citep{ale07}. A
typical simulation lasted a $\mathrm{few\!\times\!100}$ system orbital
times and resulted in $\mathrm{few\!\times\!100}$ snapshots of the
system configuration. In order to decrease the statistical errors
we ran $n_{\mathrm{sim}}\!=\!5$--$8$ simulations with different
initial conditions for each of the ($\gamma$,$Q$) models we studied.

In Fig. (\ref{f:Q1e6}) we plot the measured auto-correlation curves
in a typical simulation for $\tau\!\ll\!\tau_{\omega}$: $(\Delta E/E)(\tau)$,
$(\Delta J/J_{c})(\tau)$ and $(\Delta\mathbf{J}/J_{c})(\tau)$. The
curves reflect the joint effects of NR and RR. The coefficients $\eta_{s,v}$
and $\beta_{s,v}$ are measured by fitting the auto-correlation curves
in the coherent phase to the functions

\begin{eqnarray}
\left|\Delta J\right|/J_{c} & = & \sqrt{N}(m/M)\sqrt{\eta_{s\Lambda}^{2}\tau+\beta_{s}^{2}\tau^{2}}\,,\label{e:dJs}\\
\left|\Delta\mathbf{J}\right|/J_{c} & = & \sqrt{N}(m/M)\sqrt{\eta_{v\Lambda}^{2}\tau+\beta_{v}^{2}\tau^{2}}\,,\label{e:dJv}\end{eqnarray}
 where the two terms in the square root express the contributions
of NR and RR, respectively. Note that on short timescales, $\tau\!<\!\tau_{0}\!\equiv\!(\eta_{s\Lambda}/\beta_{s})^{2}$,
the RR auto-correlation curve rises as $\sqrt{\tau}$ due to the effect
of NR. It then rises as $\tau$ in the RR-dominated coherent phase
at times $\tau_{0}\!<\!\tau\!<\!\tau_{\omega}$, before turning over
again to a $\sqrt{\tau}$ rise in the accelerated random-walk phase
at $\tau\!>\!\tau_{\omega}$ (not shown in Fig. \ref{f:Q1e6}). The
NR and RR parameters were derived from the best 2-parameter fits of
Eqs. (\ref{e:alphaL}, \ref{e:dJs}, \ref{e:dJv}) to the data. To
control the star to star variance, we limited our analysis to time-lag
bins that sampled at least 0.75 of the stars in the simulation (\S
\ref{s:ACF}). The excellent fit of the data points to the predicted
auto-correlation curves seen in Fig. (\ref{f:Q1e6}) indicates that
RR is present and measurable.

\section{Results}

\label{s:results}

Although the auto-correlation analysis stabilizes against star to
star scatter in a single simulation, we still find a large simulation
to simulation scatter in the derived values of the coefficients. We
therefore constructed a large grid of near-Keplerian models (within
the computational time limitations), where RR should be clearly detected.

\begin{table}
\caption{\label{t:coeff}Measured NR and RR coefficients $^{a}$. }

\noindent \begin{centering}
{\scriptsize $\!\!\!\!\!\!\!$}\begin{tabular}{c@{\extracolsep{3pt}}c@{\extracolsep{1pt}}c@{\extracolsep{1pt}}c@{\extracolsep{4pt}}c@{\extracolsep{4pt}}c@{\extracolsep{4pt}}c@{\extracolsep{3pt}}c}
\hline 
\noalign{\vskip\doublerulesep}
{\scriptsize $\gamma$} & {\scriptsize $Q$} & {\scriptsize $n_{\mathrm{sim}}$} & {\scriptsize $\bar{\alpha}_{\Lambda}$} & {\scriptsize $\bar{\eta}_{s\Lambda}$} & {\scriptsize $\bar{\eta}_{v\Lambda}$} & {\scriptsize $\bar{\beta}_{s}$} & {\scriptsize $\bar{\beta}_{v}$}\tabularnewline[1pt]
\hline
\noalign{\vskip\doublerulesep}
{\scriptsize $1$} & {\scriptsize $10^{6}$} & {\scriptsize $5$} & {\scriptsize $10.60\!\pm\!0.58$} & {\scriptsize $4.26\!\pm\!0.27$} & {\scriptsize $6.56\!\pm\!0.46$} & {\scriptsize $1.15\!\pm\!0.10$} & {\scriptsize $2.00\!\pm\!0.16$}\tabularnewline[1pt]
\noalign{\vskip\doublerulesep}
{\scriptsize $1$} & {\scriptsize $10^{7}$} & {\scriptsize $6$} & {\scriptsize $10.85\!\pm\!0.91$} & {\scriptsize $4.60\!\pm\!0.34$} & {\scriptsize $6.82\!\pm\!0.42$} & {\scriptsize $1.28\!\pm\!0.13$} & {\scriptsize $2.17\!\pm\!0.22$}\tabularnewline[1pt]
\noalign{\vskip\doublerulesep}
{\scriptsize $1$} & {\scriptsize $10^{8}$} & {\scriptsize $6$} & {\scriptsize $11.57\!\pm\!1.52$} & {\scriptsize $4.17\!\pm\!0.18$} & {\scriptsize $6.18\!\pm\!0.29$} & {\scriptsize $1.06\!\pm\!0.09$} & {\scriptsize $1.95\!\pm\!0.13$}\tabularnewline[1pt]
\noalign{\vskip\doublerulesep}
{\scriptsize $1.5$} & {\scriptsize $10^{6}$} & {\scriptsize $5$} & {\scriptsize $12.20\!\pm\!1.23$} & {\scriptsize $4.03\!\pm\!0.27$} & {\scriptsize $6.40\!\pm\!0.42$} & {\scriptsize $1.03\!\pm\!0.10$} & {\scriptsize $1.76\!\pm\!0.15$}\tabularnewline[1pt]
\noalign{\vskip\doublerulesep}
{\scriptsize $1.5$} & {\scriptsize $10^{7}$} & {\scriptsize $6$} & {\scriptsize $15.63\!\pm\!4.33$} & {\scriptsize $4.17\!\pm\!0.33$} & {\scriptsize $6.92\!\pm\!0.53$} & {\scriptsize $1.02\!\pm\!0.06$} & {\scriptsize $1.83\!\pm\!0.08$}\tabularnewline[1pt]
\noalign{\vskip\doublerulesep}
{\scriptsize $1.5$} & {\scriptsize $10^{8}$} & {\scriptsize $8$} & {\scriptsize $19.12\pm6.05$} & {\scriptsize $5.09\!\pm\!0.52$} & {\scriptsize $8.43\!\pm\!1.11$} & {\scriptsize $1.16\!\pm\!0.11$} & {\scriptsize $2.35\!\pm\!0.35$}\tabularnewline[1pt]
\noalign{\vskip\doublerulesep}
{\scriptsize $1.75$} & {\scriptsize $10^{6}$} & {\scriptsize $6$} & {\scriptsize $13.00\!\pm\!1.15$} & {\scriptsize $3.95\!\pm\!0.19$} & {\scriptsize $6.12\!\pm\!0.39$} & {\scriptsize $1.03\!\pm\!0.09$} & {\scriptsize $1.93\!\pm\!0.17$}\tabularnewline[1pt]
\noalign{\vskip\doublerulesep}
{\scriptsize $1.75$} & {\scriptsize $10^{7}$} & {\scriptsize $8$} & {\scriptsize $19.61\!\pm\!3.76$} & {\scriptsize $4.05\pm0.53$} & {\scriptsize $6.11\!\pm\!.057$} & {\scriptsize $0.97\!\pm\!0.09$} & {\scriptsize $1.81\!\pm\!0.15$}\tabularnewline[1pt]
\noalign{\vskip\doublerulesep}
{\scriptsize $1.75$} & {\scriptsize $10^{8}$} & {\scriptsize $6$} & {\scriptsize $16.05\!\pm\!2.31$} & {\scriptsize $3.70\!\pm\!0.15$} & {\scriptsize $5.67\!\pm\!.017$} & {\scriptsize $0.97\!\pm\!0.07$} & {\scriptsize $1.80\!\pm\!0.09$}\tabularnewline[1pt]
\hline
\noalign{\vskip\doublerulesep}
\multicolumn{3}{l}{{\scriptsize Grand average$^{b}$}} & {\scriptsize $\left\langle \alpha\right\rangle $} & {\scriptsize $\left\langle \eta_{s}\right\rangle $} & {\scriptsize $\left\langle \eta_{v}\right\rangle $} & {\scriptsize $\left\langle \beta_{s}\right\rangle $} & {\scriptsize $\left\langle \beta_{v}\right\rangle $}\tabularnewline[1pt]
\noalign{\vskip\doublerulesep}
 &  & $56$ & {\scriptsize $3.65\!\pm\!0.28$} & {\scriptsize $1.06\!\pm\!0.03$} & {\scriptsize $1.65\!\pm\!0.05$} & {\scriptsize $1.07\!\pm\!0.03$} & {\scriptsize $1.97\!\pm\!0.07$}\tabularnewline[1pt]
\hline
\noalign{\vskip\doublerulesep}
\multicolumn{8}{l}{{\scriptsize $^{a}$ The quoted errors are the errors on the mean
(the sample rms is $\sqrt{n_{\mathrm{sim}}}$ times larger).}}\tabularnewline[1pt]
\noalign{\vskip\doublerulesep}
\multicolumn{8}{l}{{\scriptsize $^{b}$ $\left\langle \alpha\right\rangle \!=\!\left\langle \alpha_{\Lambda}/\sqrt{\ln\Lambda}\right\rangle $,
$\left\langle \eta_{s,v}\right\rangle \!=\!\left\langle \eta_{s,v\Lambda}/\sqrt{\ln\Lambda}\right\rangle $
over all simulations for $\Lambda\!=\! Q$.}}\tabularnewline[1pt]
\hline
\end{tabular}
\par\end{centering}
\end{table}

\marginpar{%
}%
\begin{figure}
\begin{centering}
\includegraphics[width=1\columnwidth]{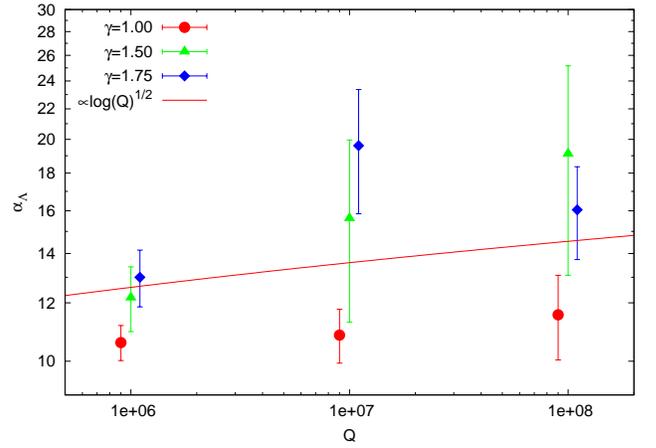}
\par\end{centering}

\caption{Measured NR energy coefficient $\alpha_{\Lambda}$ as function of
mass ratio $Q$ and for stellar density profiles with logarithmic
slopes of $\gamma\!=\!1,1.5,1.75$. A $\propto\sqrt{\ln Q}$ curve
is shown to guide the eye.}

\begin{centering}
\label{f:alphaL}
\par\end{centering}
\end{figure}

\begin{figure}
\begin{centering}
\includegraphics[width=1\columnwidth]{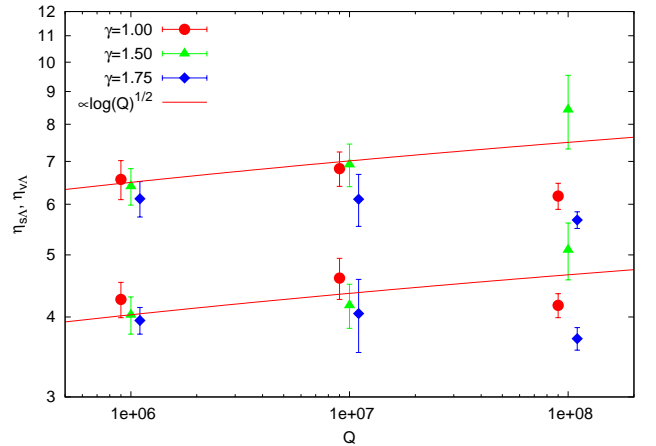}
\par\end{centering}

\caption{Same as Fig. (\ref{f:alphaL}), for the measured NR scalar angular
momentum coefficient $\eta_{s\Lambda}$ (bottom points) and the vector
angular momentum coefficients $\eta_{v\Lambda}$ (top points). }

\begin{centering}
\label{f:etasL}
\par\end{centering}
\end{figure}

\begin{figure}
\begin{centering}
\includegraphics[width=1\columnwidth]{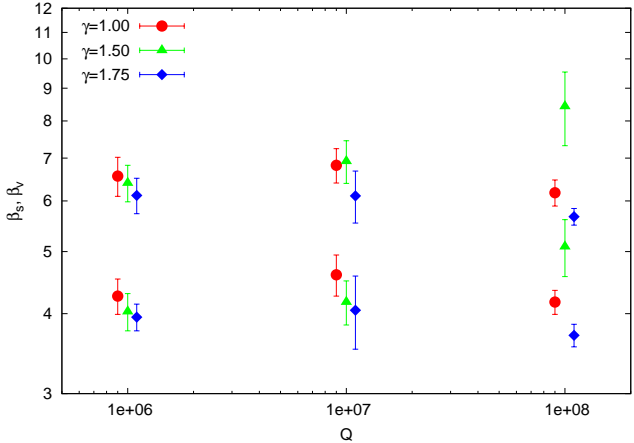}
\par\end{centering}

\caption{Same as Fig. (\ref{f:alphaL}), for the measured RR scalar angular
momentum coefficient $\beta_{s}$ (bottom points) and vector angular
momentum coefficient $\beta_{v}$ (top points). RR does not depend
on a Coulomb factor.}

\begin{centering}
\label{f:betas}
\par\end{centering}
\end{figure}

We summarize our results in table (\ref{t:coeff}) and in Figs. (\ref{f:alphaL},
\ref{f:etasL}, \ref{f:betas}). The coefficients $\alpha_{\Lambda}$
and $\eta_{s,v\Lambda}$ do not directly express the intrinsic properties
of NR, since they should vary as $\sqrt{\textrm{ln}\Lambda}\!\simeq\!\sqrt{\ln Q}\!\simeq\!4(1\!\pm\!0.07)$
over the $Q$-range of our models. This small fractional difference
and the relatively low statistics of our simulation suite may explain
why we do not detect a clear $Q$-dependence in these quantities.
Since we do not see a clear $\gamma$ or $Q$ dependence, we adopt
as the best fit estimates of the values of $\alpha$, $\eta_{s,v}$
and $\beta_{s,v}$ their grand average over all the simulations (table
\ref{t:coeff}). The measured ratio $\left\langle \beta_{v}/\beta_{s}\right\rangle \!=\!1.85$
is consistent with the predicted ratio of $\sqrt{3}\!\simeq\!1.73$
(\S \ref{sss:sv}). 

RT96 derived from their simulations smaller mean values for $\beta_{s}$
($0.53$, as compared to $1.07$ here) and for $\alpha_{\Lambda}$
and $\eta_{s\Lambda}$ ($3.09$ and $1.37$ as compared to $14.76$
and $4.25$ here). At least part of the difference in $\alpha_{\Lambda}$
and $\eta_{s\Lambda}$ can be traced to their use of a softening length
$\epsilon=10^{-2}R\!\gg\! Gm/v^{2}\!\sim\!(m/M)R$ in the calculation
of the gravitational force. This decreases the effective value of
the Coulomb factor, since $\Lambda\!\sim\! R/\max(\epsilon,Gm/v^{2})\!=\!10^{2}$
and so $\sqrt{\ln\Lambda}\!\simeq\!2.1$, about twice as small as
in our simulations. Indeed, RT96 noted that decreasing the softening
length to $\varepsilon\!=\!10^{-4}R$ led to an increased value of
$\alpha_{\Lambda}\!=\!5.5\pm0.2$. 

\begin{figure}[t]
\begin{centering}
\includegraphics[width=1\columnwidth]{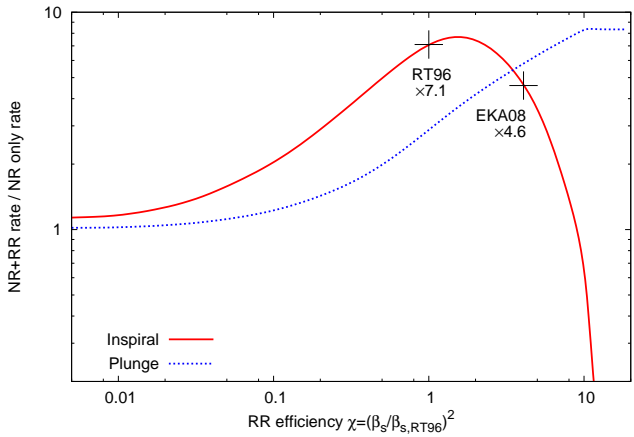}
\par\end{centering}

\caption{\label{f:EMRI}The change in the EMRI and plunge rates relative to
that predicted assuming only NR, as function of the RR efficiency
$\chi$ (Adapted from \citealt{hop+06a}, Fig. 5). The RT96 estimate
of $\beta_{s}$ ($\chi\!=\!1$) predicts an increase of $\times7.1$
in the EMRI rate due to RR, close to the maximum. Our (EKA08) new
measured efficient RR ($\chi\!=\!4.1$) predicts a higher plunge rate
and thus a smaller increase of $\times4.6$ in the EMRI rate.}

\end{figure}

It is difficult to trace the specific reason for the discrepancy between
our best estimate value for $\beta_{s}$ and that derived by RT96,
given the many differences in both the simulations and the methods
of analysis. Our statistics are better due to the larger number of
simulations and more efficient use of the data, and the analysis here
is more rigorous. We therefore briefly consider the implications of
this revised value of $\beta_{s}$ for EMRI rates. \citet{hop+06a}
parametrized the RR efficiency by a factor $\chi\!=\!\left[\beta_{s}/\beta_{s,\mathrm{RT}96}\right]^{2}$,
and derived the $\chi$ dependence of the branching ratio of the EMRI
and infall (plunge) rates (Fig. \ref{f:EMRI}). The RT96 value $\chi\!=\!1$
happens to lie very close to the maximum of the RR-accelerated EMRI
rate. We find here $\beta_{s}/\beta_{s,\mathrm{RT}96}\!\sim\!2$,
which corresponds to a factor $\sim\!5$ increase in the EMRI rate
compared to that estimated for NR only, but is a factor $\sim\!1.5$
smaller than implied by the RT96 value, because the higher RR efficiency
leads to a higher plunge rate at the expense of the inspiral rate.

\section{Discussion and summary}

\label{s:conclusions}

We characterized and measured the mean efficiency coefficients of
NR ($\alpha_{\Lambda},\eta_{s,\Lambda},\eta_{v,\Lambda}$) and RR
($\beta_{s},\beta_{v}$) in Newtonian $N$-body simulations of isotropic,
thermal, near-Keplerian stellar cusps around a MBH. We derived a simple
analytical form for the rms RR auto-correlation curves of $\mathbf{J}$
and $J$, and showed that the two are simply proportional to each
other. We then measured these coefficients in a large suite of small
scale $N$-body simulations with different stellar density distributions
and MBH/star mass ratios. We found no statistically significant trends
in the values of these coefficients as function of the system properties.
This may require better statistics. 

Our measured RR efficiency suggests that RR increases the EMRI rate
by a factor of $\sim\!5$ above what is predicted for NR only. This
estimate of RR efficiency is consistent with that suggested by the
analysis of the dynamical properties of the different stellar populations
in the Galactic Center \citep{hop+06a}. However, this conclusion
is still preliminary, since several important open issues remain,
which should be addressed by larger scale simulations. These include
(1) The dependence of the RR coefficients on the orbit of the test
star, for example its eccentricity \citep{gur+07}. We find that $N\!\sim\!200$
is not enough for reliable statistics on sub-samples within given
\emph{E} or \emph{J}-bins. The eccentricity dependence of RR is particularly
relevant for the supply rate of stars to the MBH from $J\!\rightarrow\!0$
orbits (the loss-cone refilling problem). (2) The effects of a stellar
mass spectrum. This will likely affect RR by decreasing the RR timescale,
and changing the stellar density distribution through strong mass
segregation (\citealt{ale07}; Alexander \& Hopman 2008, in prep.).
(3) The robustness of RR against perturbations from the larger non-Keplerian
stellar system in which the inner near-Keplerian region of interest
is embedded. (4) The role of post-Newtonian effects in RR, such as
General Relativistic precession and GW emission. These are expected
to play a key role in enabling inspiral by quenching RR just as the
compact remnant enters the EMRI phase, and in regulating the GW inspiral
rate \citep{hop+06a}.

\acknowledgements{TA is supported by ISF grant 928/06, ERC Starting Grant 202996-2
and a New Faculty grant by Sir H. Djangoly, CBE, of London, UK.}

\bibliographystyle{apj}

\end{document}